\documentclass[final,3p,times]{elsarticle}

\usepackage{amsmath}
\usepackage{hyperref}
\usepackage{natbib}

\usepackage{amssymb}

\biboptions{sort&compress}

\newcommand{\ba}{\begin{eqnarray}}
\newcommand{\ea}[1]{\label{#1} \end{eqnarray} }

\begin{document}

\begin{frontmatter}



\title{\bf \Large A proposal of a concept for a balanced internal combustion engine configuration}


\author[label1]{Istv\'an Gere \corref{cor1}},
\author[label2]{Ovidiu-Florin Botoș}
\author[label2]{Liviu Ancuțescu}

\address[label1]{Physics Department, Babe\c{s}-Bolyai University, Cluj-Napoca, RO-400347, Romania}
\address[label2]{Cluj-Napoca, RO-400347, Romania}

\cortext[cor1]{Corresponding author: istvan.gere$@$ubbcluj.ro}

\begin{abstract}
Conventional internal combustion engines (ICEs) inherently suffer from imbalances due to the reciprocating motion of their components, leading to vibrations, wear, and mechanical inefficiencies. In this paper, we propose a novel engine configuration based on four interconnected hypocycloidal straight-line mechanisms, designed to eliminate both translational force imbalances and torque fluctuations. We present the mechanical concept as a physics problem, followed by theoretical calculations showing that the proposed system results in zero net force and torque due to its symmetric design. An experimental prototype was constructed using CAD modeling and manufactured from acrylic glass. Measurements performed with an accelerometer confirm the theoretical predictions: the engine remains stable during operation, and imbalances arise only when the piston synchrony is disrupted. Our results demonstrate that such a configuration offers a mechanically balanced alternative to conventional ICE layouts, with potential applications in vibration-free piston-based engine designs.\end{abstract}




\end{frontmatter}







\section{Introduction}
 Internal combustion engines (ICE as short) are a class of rotary machines that transform the reciprocating motion of pistons into rotational motion, which can then be used for various purposes (such as driving the wheels of a vehicle, generating electricity etc.). A longstanding problem in the design of such rotary machines, is the issue of dealing with imbalances that are present due to the physical layout. Ever since the last century, considerable scientific work has been dedicated to mitigation solutions of these problems\cite{book1}.The vibrations resulting from the unbalanced components of such machines performing reciprocating and rotating motion are known to put a prescribed reliability and lifetime limit on them, as these issues cannot be avoided in classical ICE designs\cite{review}.Along other considerations, one of the sources of the diverse ICE layouts that exist can be attributed to the attempt of canceling out the inertial forces that are the cause of these vibrations\cite{layout_review}. In this regard, in-line engines are considered to be unbalanced, while the flat boxer engine is considered to be balanced from the point of view of the forces acting upon the pistons\cite{flat_boxer}, but torque resonances on the crankshaft are still present\cite{boxer_fail}. Other considerations in reducing ICE vibrations is driving comfort\cite{vibration_conform,vibration_conform_2} and possibly manufacturing advantages due to simpler engine mounts. Interest in the problem of balancing engines and engine components are also found in the literature\cite{balanced_engine_concpt, balanced_cramshaft}
 \par
In this article we present a simple concept of an engine layout that is balanced from the point of view of the inertial forces resulting from the reciprocating motion of the pistons, as well from the point of view of the sum of the resulting torques. To establish this configuration, we went back to the hypocycloidal straight-line mechanism designs from the 19th century. Using four of these mechanisms, it is possible to create a layout with the above-mentioned properties. In Section \ref{sec:problem-of-balanced-engine} we present our engine layout as a physics problem, in Section \ref{sec:theoretical-hypo} we show with theoretical calculations that such a layout is balanced while in Section \ref{sec:experimental} we present an experimental realisation of such a system along with measurements that validate our theory. In the last Section \ref{sec:conclusion} we draw our conclusions with  outlooks to the additional benefits of such a layout.
\newline


\section{The problem of a balanced engine layout from a physics point of view
\label{sec:problem-of-balanced-engine}}
\newcommand{\be}{ \begin{equation} }
\newcommand{\ee}[1]{\label{#1} \end{equation} }

 It can be intuitively seen that having a single piston configuration in a cylinder, along with a crankshaft, will result in inertial forces acting axially due to the reciprocating movement of the piston, and radially due to the left-right movement of the connecting rod around the crankshaft. A resultant torque countering the direction of the rotation of the crankshaft will also be present on the whole assembly of the engine (which is usually canceled by the rotating movement of a counterweight). As we can see, such an engine will be characterized by vibrations and wear (the characteristic oval wear of the cylinders due to the radial forces\cite{oval_wear}). MMultiple piston configurations such as the flat and boxer engine may reduce a part of the resulting vibrations from the reciprocating movement\cite{flat_boxer} at the cost of increased manufacturing and engine complexity. We may note that the problem of the unpaired torques are still present in the boxer configurations which may result in critical failure\cite{boxer_fail}.
\par
As such in the simplest terms, the search for a balanced engine configuration may be considered a physics exercise. Different engine layouts may be considered where the resultant inertial forces cancel each other out along with the resulting torques leading to a state of balance. We found an elegant engine layout consisting of four pistons performing translational motion connected to hypocycloidal straight-line mechanisms transforming it into rotational motion. We connected these four rotating mechanisms with gears and concluded that this layout leads to a balanced state. In the next section we will present this device.

\section{The theoretical considerations of an ICE configuration with four hypocycloidal components}
\label{sec:theoretical-hypo}

\begin{figure}[!h]
    \centering
    \includegraphics[width=.9\textwidth]{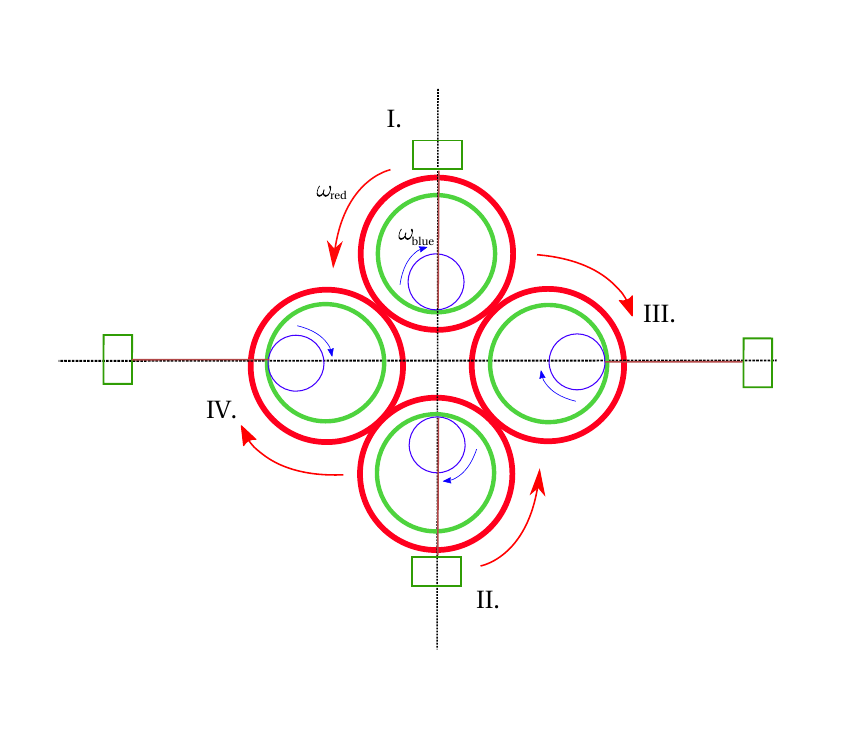}
    \caption{A sketch of the balanced hypocycloidal engine layout, with circles that transfer the rotational motion through coupling. The dark green squares represent the pistons, linked with a brown colored rod to the blue disks. The blue disks perform the hypocycloidal motion inside the stationary light green colored circle. The blue disks are then connected with a shaft to the red disks. The red disks are synchronized as meshing gears adding up the motion of the subcomponents.}
    \label{fig:samplesetup}
\end{figure}

As a diagrammatic abstraction, we are using disks rotating with perfect rolling (in the real acrylic model these are meshing gears). We consider an engine where four pistons (colored dark green in Fig. ~\ref{fig:samplesetup}) are each linked to one of four distinct, smaller disks (colored blue) which, in turn, are connected with a shaft to a larger disk (colored red). In this setup the blue disks will perform a hypocycloidal motion within the stationary light green colored circle representing the internal gear (this circle does not move in relation with the engine assembly), thus transforming the linear motion to a rotational one. The larger red disks are synchronized as meshing gears as presented in Fig. ~\ref{fig:samplesetup}.

\subsection*{The motion of the pistons}

We deduce the resulting force $F_r$  of a coupled four body system moving with the supposition that the masses of the bodies are equal:

\begin{equation}
   m_I = m_{II} = m_{III} = m_{IV} = m
   \label{Eq:equation1} 
\end{equation}

We consider that the pistons exhibit a linear harmonic motion as described in Eq. \ref{linear_harm_motion}.

\begin{equation}
    \begin{array}{l}
    a(t) = - \omega_p^2 x(t)
    \\
    x(t) = A \cdot cos(\omega_p t + \phi)
    \\
    F = m \cdot a(t)
    \\
    F(t) = -m\omega_p^{2}x(t)
    \end{array}
    \label{linear_harm_motion}
\end{equation}

Here $\arrowvert x_0 \arrowvert = A $  notes the amplitudes of the motion. The following equations denote the phases of the pistons noted with roman numerals in Figure \ref{fig:samplesetup}:
\begin{equation}
    \begin{array}{l}
    I \rightarrow \phi_{I} = 3\pi/2
    \\
    II \rightarrow \phi_{II} = \pi/2
    \\
    III \rightarrow \phi_{III} = 0
    \\
    IV \rightarrow \phi_{IV} = \pi
    \end{array}
    \label{eq333}
\end{equation}

With these phase differences we can write the resulting forces action caused by the reciprocating motion of the pistons as $F_r$.

\begin{equation}
    \begin{array}{l}
    F_r = F_I + F_{II} + F_{III} + F_{IV} =
    \\
    = m(\omega_{p_{I}} ^2 x(t) +\omega_{p_{II}} ^2x(t) +\omega_{p_{III}} ^2 x(t) + \omega_{p_{IV}} ^2 x(t))
    \end{array}
    \label{eq444}
\end{equation}

From the phase differences in Eq. \ref{eq333} and the equation of the resulting force in Eq. \ref{eq444}  we observe that the phase difference between the piston pair $F_I$, $F_{II}$ and pair $F_{III}$, $F_{IV}$ is $\pi$, as such we can rewrite Eq. \ref{eq444} as:
\begin{equation}
 F_r = m(\omega_p^2 x(t) - \omega_p^2x(t) x +\omega_p^2 x(t) - \omega_p^2 x(t)) = 0
\end{equation}
As the system is coupled, the angular frequencies $\omega$ are equal, and only their sign changes due to the phase differences.

\subsection*{The resulting torque of the system}

The blue disks performing the hypocycloidal motion, to which the pistons are linked have the angular frequency of $\omega_{blue}= \omega_{red}/2$.During a full rotation of the blue disks around their center the red disks perform half of a rotation. The torque resulting from the rotation of the red disks can be written as Eq. \ref{eqqq_torque}.
\begin{equation}
 \tau_{red} = I_{red}  \alpha_{red}
 \label{eqqq_torque}
\end{equation}

In Eq. \ref{eqqq_torque} $I$ is moment of inertia of the red disks and $\alpha_{red} = d \omega_{red} / dt$ is the angular acceleration. The resultant torque from the rotation of the red disks can be calculated from the sum of the torques for the individual disks as presented in Eq. \ref{eqqq_torque_summ}.

\begin{equation}
    \begin{array}{l}
    \tau_{R} = \tau_{red_{I}} + \tau_{red_{II}} + \tau_{red_{III}} + \tau_{red_{IV}}
    \end{array}
    \label{eqqq_torque_summ}
\end{equation}

Eq. \ref{eqqq_torque_summ} can be expanded and simplified as the red disks are identical, and their moment of inertia is equal $I_{red} = I_{I} = I_{II} = I_{III} = I_{IV}$:

\begin{equation}
    \begin{array}{l}
    \tau_{R}
    = I_{red} \frac{d \omega_{red_{I}}}{dt} + I_{red} \frac{d \omega_{red_{II}}}{dt} +I_{red} \frac{d \omega_{red_{III}}}{dt}  + I_{red} \frac{d \omega_{red_{IV}}}{dt}
    \end{array}
\end{equation}

The resultant torque $\tau_R =0$ s equal to zero due to the symmetry of the system as the rotation of the red disks are of the opposite direction.
\par
The blue disks perform two rotations, one around their own center as the axis, and one around the central axis of the red disks. In both conditions the resulting torques are zero. We note that if the system had only two coupled hypocycloidal mechanisms, the resultant torque would not equate to zero, as the inner blue disks’ rotation around the axis of the red disks are in the same direction, meaning that a resultant torque would be present, and that will spin the entire assembly.

\section{An experimental realisation of the balanced hypocycloidal concept}
\label{sec:experimental}

\begin{figure}[!h]
    \centering
    \includegraphics[width=.4\textwidth]{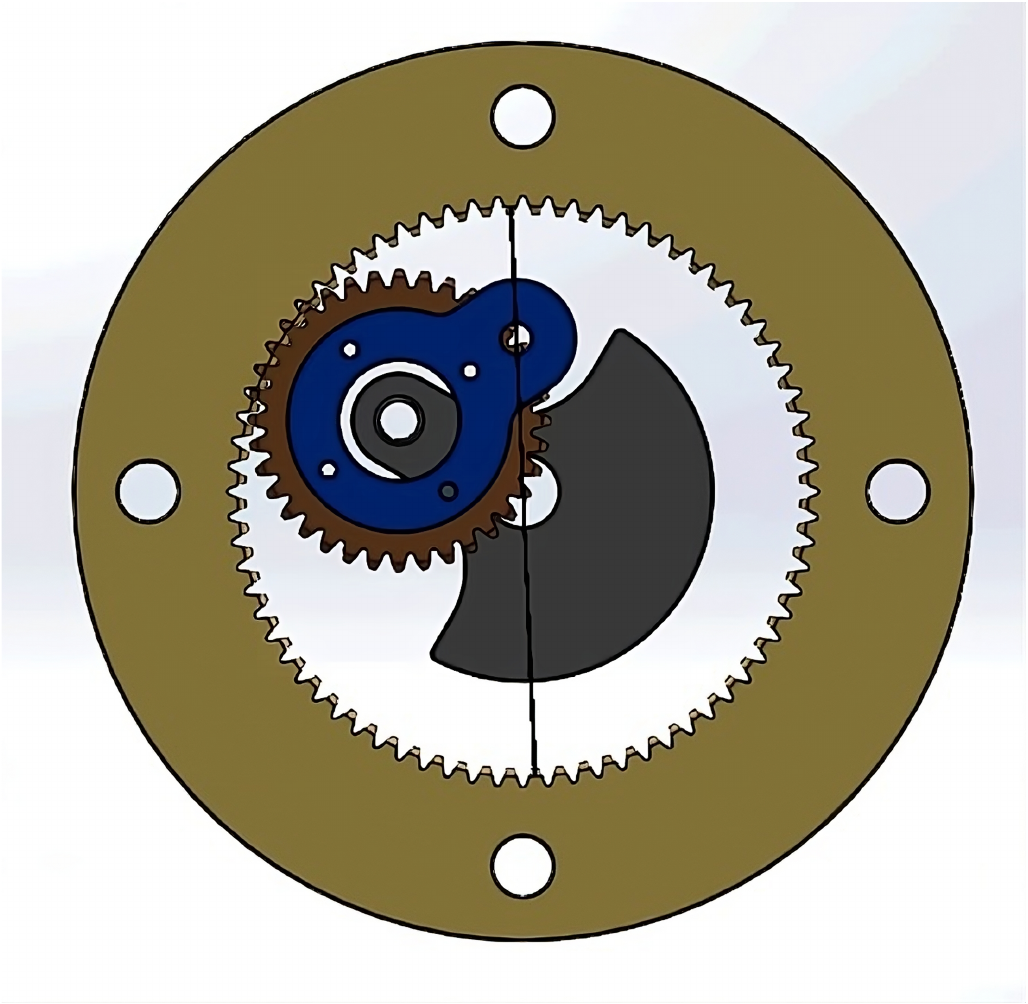}
    \caption{he CAD model of a single hypocycloidal component from the engine block.}
    \label{fig:hypo_single}
\end{figure}

To test the balanced engine concept described in the previous chapter, we created a prototype of the engine. The individual hypocycloidal components of the engine are presented in Figure \ref{fig:hypo_single}

\begin{figure}[!h]
    \centering
    \includegraphics[width=.9\textwidth]{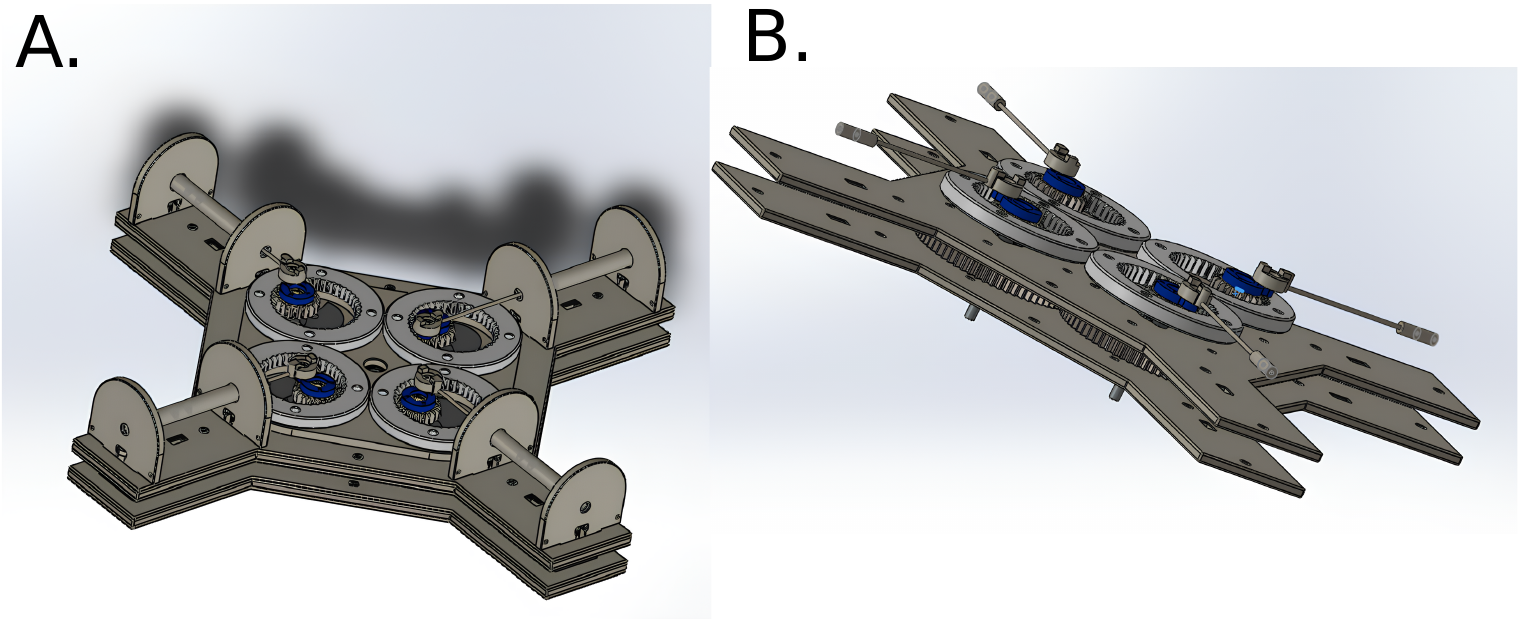}
    \caption{\textbf{A.} The CAD model of the prototype engine block with four hypocycloidal components in a cross layout. \textbf{B.} he CAD model from another angle, presenting the underlying gears that connect the hypocycloidal components.}
    \label{fig:hypo_ans}
\end{figure}

The whole engine is made out of four of these hypocycloidal components in a cross like layout on the top layer, while the second layer these components are connected with additional synchronization gears as presented in Fig. \ref{fig:hypo_ans}.

\begin{figure}[!h]
    \centering
    \includegraphics[width=.9\textwidth]{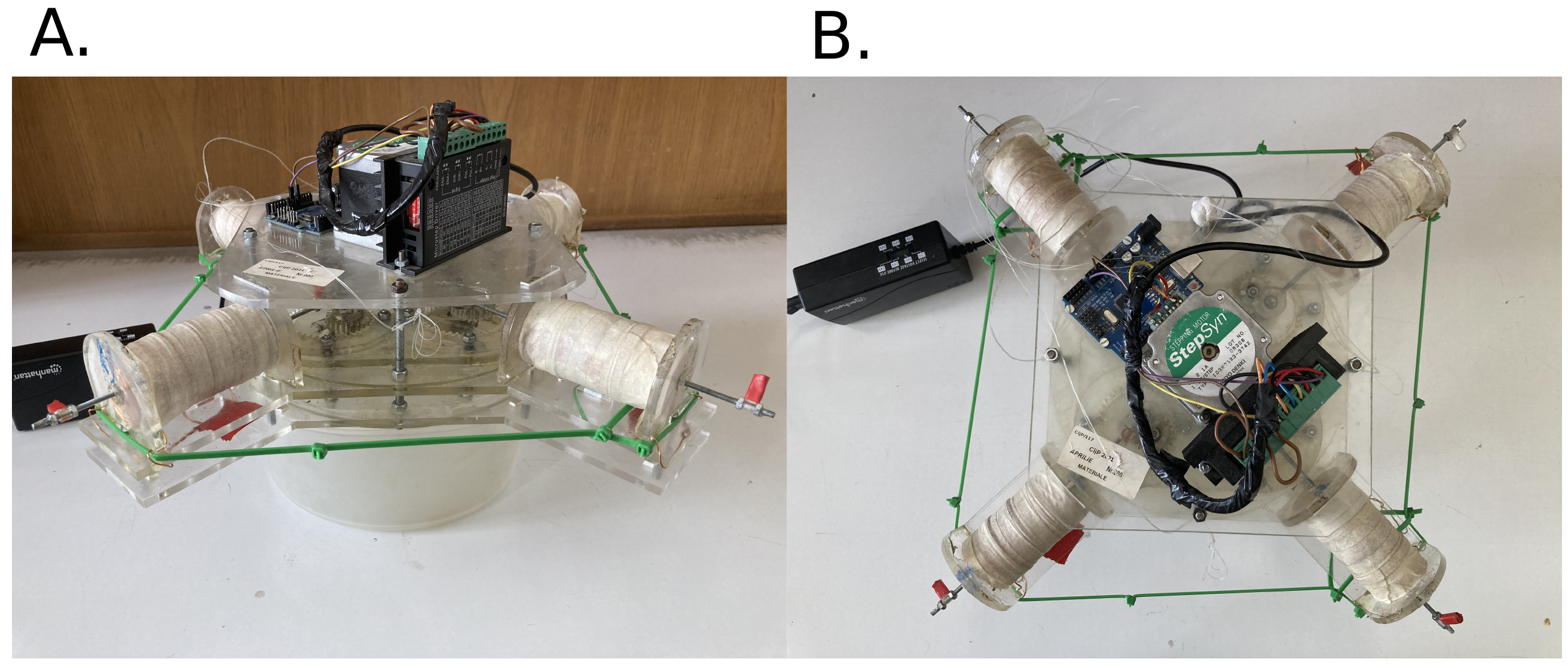}
    \caption{\textbf{A.} A side view of the model. \textbf{B.} A top view of the model.}
    \label{fig:hypo_model}
\end{figure}

The physical model of the engine was fabricated from acrylic. The different layers of the engine were mounted together, and the hypocycloidal components were aligned and synchronized in order to ensure smooth movement. Three additional gears were added in the third lowest part of the engine that are connected to an Arduino-controlled motor which drives the whole assembly. We present pictures of the model in Figure \ref{fig:hypo_model}.We used this model to perform our measurements regarding the physical stability of the mechanism.

\newpage
\subsubsection*{Experimental measurements on the concept}

\begin{figure}[!h]
    \centering
    \includegraphics[width=.8\textwidth]{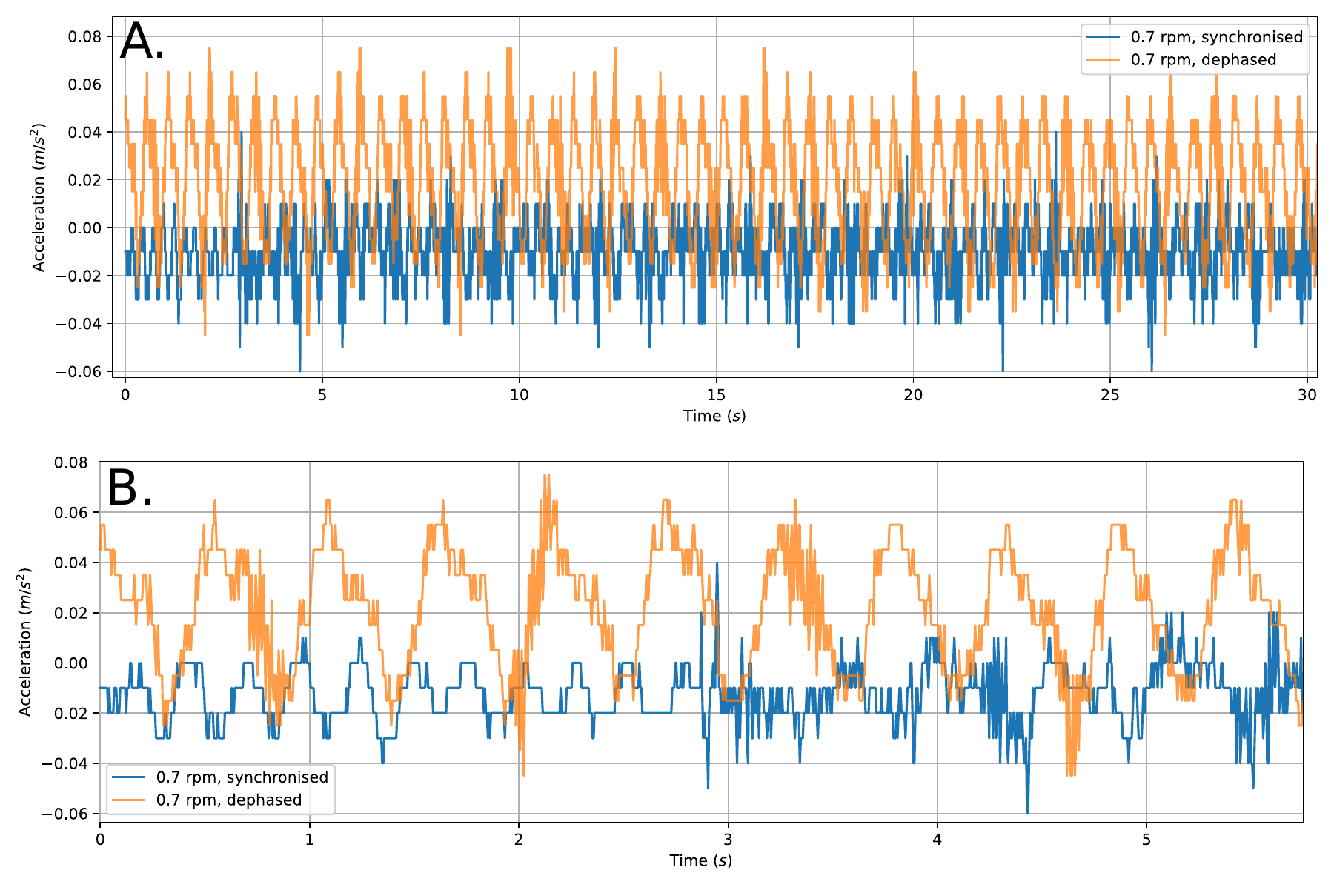}
    \caption{The accelerations detected on the mechanism when it was rotated with 0.7 rpm, in the syncronized and de-phased cases presented on: \textbf{A.} A larger timeframe. \textbf{B.} A zoomed-in timeframe.}
    \label{fig:hypo_acceleration}
\end{figure}

The motor assembly was then suspended with plastic fishing wires attached to four symmetric points on the motor assembly so it was able to freely move if any asymmetric forces or torques were present. An accelerometer was placed on the engine. Multiple tests were run with different speeds, while no motion of the engine were observed in synchronized configuration. Then one of the pistons was de-phased with 180 degrees in relation to its initial position, to break the symmetry of the mechanism. The suspended engine in this situation starts a periodic motion as anticipated. The resulting accelerations are presented in Figure \ref{fig:hypo_acceleration}.

\newpage
\section{Conclusions}
\label{sec:conclusion}
In this article we present a novel approach to the problem of the conventional unbalanced internal combustion engine layout. It is known that the classical ICE designs are unbalanced, caused by the movement of the pistons and the torques present in the engine. This in turn leads to wear, driving discomfort and other problems (ex. any unwanted vibration represents lost energy). By introducing a novel hypocycloidal based engine layout with four pistons, we proved mathematically that such a layout is completely balanced. For easier presentation of the concept, a model was built from acrylic. We suspended the model with wires, so it was able to freely move and we placed an accelerometer on it, to show that there is almost zero shaking compared to a state in which one of the pistons was de-phased with 180 degrees from its initial position. Our novel approach can be applied in ICEs allowing smooth running operation, solving the problem of mechanical balance. If built as an ICE it may allow for ceramic components to be used as there are no lateral forces to make the ceramics exceed their mechanical limits.

\section*{Acknowledgements}
The work of István Gere was supported by a grant of the Ministry of Research, Innovation and Digitization, CNCS/CCCDI - UEFISCDI, project number PN-IV-P8-8.1-PRE-HE-ORG-2023-0118, within PNCDI IV. The mention of Bosch as an affiliation refers solely to the author's current employment status and does not imply any contribution, support, endorsement, intellectual property involvement, or material assistance from Bosch regarding the development of this work. All concepts, experimental results, designs, and conclusions presented herein are solely the result of the independent research and personal initiative of Ovidiu-Florin Botoș and Liviu Ancuțescu.

\section*{Author contributions statement}
O.F.B. has conceptualized the idea of the balanced configuration based on the classical hypocycloidal straight-line mechanism. The mathematical and physical aspects of the balanced configuration were investigated by I.G.. O.F.B. and L.A.. created the CAD and physical model of the balanced engine configuration. Measurements on the model were performed by I.G. and O.F.B.. Visualizations were created by O.F.B., L.A. and I.G. The manuscript has been written and edited by I.G. and O.F.B.. All authors contributed.





\end{document}